\documentclass{article}

\usepackage{hyperref}
\usepackage{graphicx}
\usepackage{subfig}

\begin{document}



\title{Dynamic Transposition of Melodic Sequences on Digital Devices}

\author{
A.V. Smirnov,
andrei.v.smirnov@gmail.com. 
}

\maketitle 

\sffamily\mdseries

\begin{abstract}
A method is proposed which enables one to produce musical compositions by using
transposition in place of harmonic progression.  A transposition scale is
introduced to provide a set of intervals commensurate with the musical scale,
such as chromatic or just intonation scales.  A sequence of intervals selected
from the transposition scale is used to shift instrument frequency at
predefined times during the composition which serves as a harmonic sequence of
a composition.  A transposition sequence constructed in such a way can be
extended to a hierarchy of sequences.  The fundamental sound frequency of an
instrument is obtained as a product of the base frequency, instrument key
factor, and a cumulative product of respective factors from all the harmonic
sequences.  The multiplication factors are selected from subsets of rational
numbers, which form instrument scales and transposition scales of different
levels.  Each harmonic sequence can be related to its own transposition scale,
or a single scale can be used for all levels.  When composing for an orchestra
of instruments, harmonic sequences and instrument scales can be assigned
independently to each musical instrument.  The method solves the problem of
using just intonation scale across multiple octaves as well as simplifies
writing of instrument scores.

\end{abstract}

%

\section{Introduction}

Repetition is one of the main elements of music.  Indeed, a common technique
used in composition is a repetition of a melodic sequence in different keys.
The change of keys introduces another important element of music, that of
gradual change (harmonic/chord progression or sequence), which removes the
monotony of a simple repetition. In principle, the change of key in harmonic
progression amounts to a frequency shift and can be done by transposition.  For
example, the technique of using barre chords when playing a guitar does exactly
that, and this greatly enhances the versatility of the instrument.  Another example is a tuning slide of a trombone.

To accomplish the same on a key-based instrument one can supply such instrument
with extra transposition keys. This set of keys can be used during the
performance to transpose the instrument each time the change of harmonic key
needs to be done within the composition. In this case the melody can be played
as if it always stays in the same key, which simplifies playing technique.
Also, a number of keys in an octave of such instrument can be reduced to only
the keys which form harmonic intervals, since most melodic sequences tend to
comply to these intervals. For example, in chromatic scale one would use 8
tones, including major and minor triads, instead of 12 semitones. 

For example, the octave in a conventional chromatic scale is sub-divided into
12 intervals (semitones). Thus, in an instrument with a range of three octaves
one would need 36 buttons (or instrument keys) to play all notes. However, one
can achieve essentially the same by using six buttons for playing the notes and
other six buttons to do the transposition.  This will result in a three-fold
reduction in the number of instrument keys (or buttons), and the reduction will
be greater if more octaves are involved.  Moreover, since most melodies when
played in the same musical key use only a small subset of 12 semitones, one can
assign the scale tones to the play buttons and use the transposition buttons
for switching the musical keys.  For example, one could use only eight tones of
the chromatic scale, including major and minor triads, instead of 12 semitones.
This way one will only engage the transposition keys when changing the musical
key, therefore simplifying the performance.  

There is another important advantage of such an instrument. Unlike a
conventional keyboard or a guitar, it can be made to play in scales other than
chromatic scale. One can easily assign the frequencies of a Pythagorean or just
intonation scale to the keys of such instrument.  This is because the
uniformity of musical intervals within the octave is no longer required, since
the scale played by the instrument keys is easily transposed to different
musical keys.

This kind of transposition cannot be easily accomplished on an acoustic
instrument. For example, the barre technique on a guitar will fail because
transposition in a just intonation scale will require simultaneous re-tuning of
all the strings.  This is because just intervals are not uniform across the
octave as in chromatic scale.  On the other hand, an electronic instrument or a
computer sequencing application can easily accommodate such technique which we
shall refer to as {\em dynamic transposition} and which is the main subject of
this work.


\section{The Choice of Scale}

\begin{figure}[h]
\centering
\includegraphics[width=4.8in]{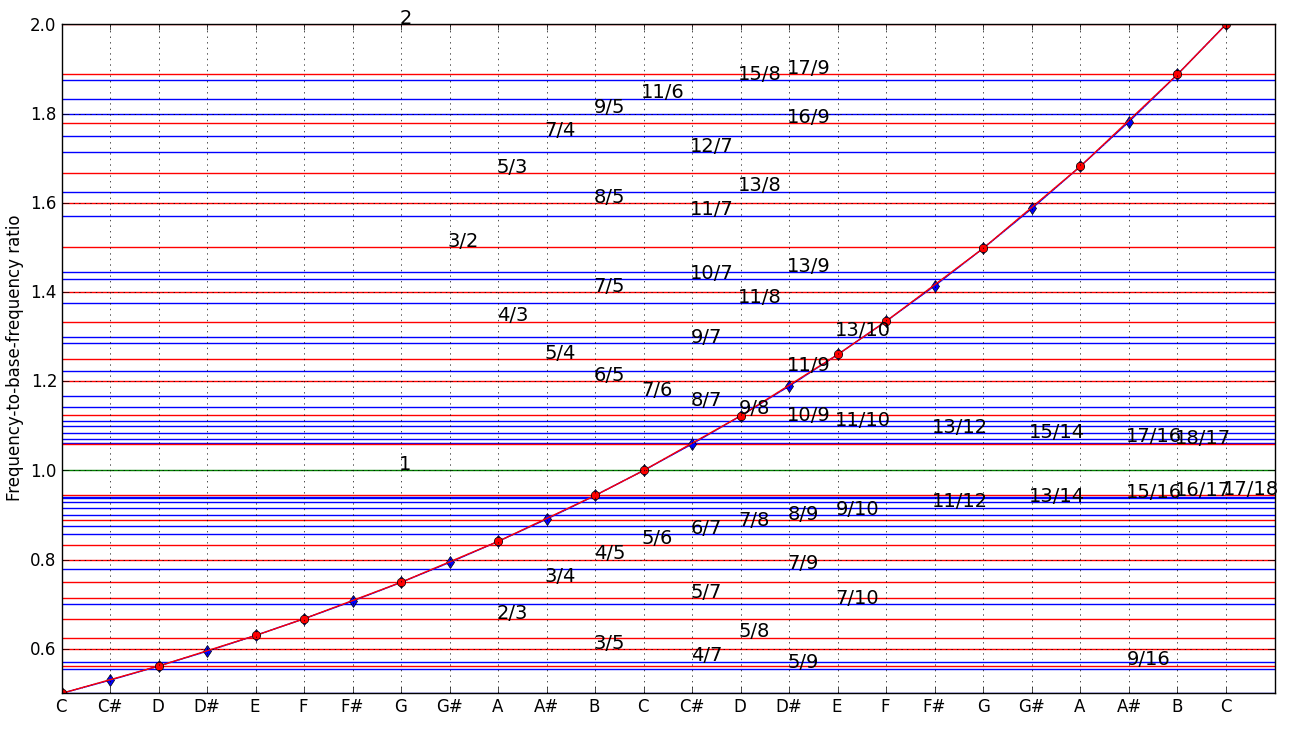}
\caption{\label{chromatic}~Frequencies of chromatic scale
}
\vspace{5mm}
\parbox{4in}{
Number above each line added to one is equal to the scaling factor used to obtain the
frequency associated with that line. Red points correspond to the C-major scale.
}
\end{figure}

One limitation of a standard chromatic scale used in music
today is the inexact representation of harmonic tones, as derived from
the physics of resonating strings \cite{WdPM1980}. Harmonic frequencies
differ by rational multipliers,  for example 2, 1/2,
3/2, 3/4, 4/3. Sounds combined of such frequencies are usually pleasant
to the ear.  This could be related to resonances caused by harmonic
frequencies inside the ear, or some more complex phenomena inside the
brain \cite{LevYBM2007,CkMCCS2001}.  
Indeed, earlier instruments were based on resonating strings, and were tuned to
follow these ratios which is reflected, for example, in Pythagorean scale
\cite{BnMTP2003}. Here we will refer to a {\em rational scale} as a just
intonation scale with intervals perceived as consonant.  A significant
limitation of a rational scale is that it is not {\em transpositionally invariant} in
a sense that the same chords played in different keys will not sound the same
(this definition differs somewhat from \cite{MlIFATC07}).  This is because the
intervals in a scale like 1, 6/5, 5/4, 4/3, 3/2, are not equal whereas in
chromatic scale they are equally spaced on a logarithmic scale as is
illustrated in Fig.\ref{chromatic}. This makes it impossible to play the same
melody in different keys on a rational scale.  Thus, a scale derived from
rational numbers can not be easily applied to a key-based instrument, such as a
piano, or an organ.  Elaborate tuning systems and scales combined with novel
keyboard concepts have been devised to overcome this limitation
\cite{Fokker67,Fokker69,Fontville91,MlIFATC07}, but these techniques 
added
to the complexity of the keyboard and required extra technical skills from a performer.  In a addition to that a variety of sound
synthesis methods as well as spectral music techniques \cite{SpectralMusic01}
and frequency modulation synthesis \cite{FMsynth73} have been developed, which
focus more on acoustical quality of sound rather than on composition techniques and melodic
sequencing, and are outside of the scope of this work.

A number of composers and performers experimented in just intonation scale,
most notably, Harry Partch, James Tenney, and Ben Johnston. In particular,
Harry Partch \cite{HarryParth} did early experiments with tunings based on a larger number of
unequal tones in the octave following just intonation and produced 
instruments which could play in this scale. James Tenney started one of the early and
successful experimentations with electronic and computer generated music, in which
he resorted to more harmoneous scales, and improved the
acoustic quality of sound.  Another approach \cite{BenJonson06} is
based on extending the musical notation by introducing more intervals in line
with just intonation scale. This work demonstrated the advantages and
drawbacks of musical instruments and composition techniques based on 
such scales. However, the methods mentioned above did not rely on transposition as means
of replacing harmonic progression in a sense the current work suggests.

In addition to that, a number of novel tuning methods have
been researched and proposed over the years \cite{CarlosTAC87}, including the
asymmetric division of octave \cite{SetharesRTT92} and adjusting the instrument
tuning to timbre \cite{SetharesTTSS2005}.  
However these techniques either
increase the complexity of an instrument or like chromatic scale, they compromise
the purity of intervals derived from the rational scale. 

Indeed, frequencies produced by chromatic scale no longer differ by rational
multipliers, but rather by transcendental numbers resulting from the
logarithmic operation as is evident from Fig.\ref{chromatic}. This is the
reason why chords played on an acoustic piano sound more pleasantly than those
played on an electronic keyboard, since the strings of an acoustic instrument
can self-adjust their frequencies due to the resonance effect. This makes them
lapse to more pleasant rational intervals \cite{CarlosTAC87}.

Thus, despite the advantages of just intonation and other rational scales, various technical problems 
prevented these techniques from gaining popularity.
One notable exception is a recent technique of program controlled tuning
\cite{Hermode11}, which actually enables one to play in just
intonation scale on electronic keyboards. However, this technique only corrects
musical tuning when chords are played, i.e. several notes played simultaneously.
It does not simplify music sequencing by taking advantage of transposition, neither does it attempt to
simplify the instrument by and reducing the
cardinality of the octave as discussed below. These simplifications constitute the essence of
dynamic transposition technique which is discussed next.

\section{Dynamic Transposition}

There are two main techniques for shifting sound frequency in music: modulation
and transposition.  Modulation has been mainly used for sound effects, or as an
occasional semi-tone shift in a composition.  A wider frequency shift was
mainly limited to octave shift in some electronic keyboards. Shifting of
instrument's frequency to a different key is commonly done with transposition.
However, this shift is usually made permanent for the duration of the
performance. To author's knowledge neither modulation nor transposition have
been widely used in place of harmonic progression, that is, to change current
key in a composition.  Nevertheless, such application will enable the usage of
rational scales, simplify a keyboard instrument, as well as reduce the number
of notes in an octave \cite{SmirnovPat2013,Tectral15}. In particular it can simplify learning of of musical scales and enable easy to play instruments.

Indeed, if we consider the number of tones in an octave, or cardinality, then
with the proposed technique it can be reduced to only the most harmonic musical
intervals, such as those in a rational scale.  This is because the number of
keys in the octave can be made equal to the number of tones in the scale.  For
example, instead of selecting 7 keys for a chromatic scale out of 12 semitones
in an octave, one is left with just 7 keys corresponding to 7 tones and no need
for selection.  And for simpler scales this number can be further reduced.
Now, instead of a multitude of tonic combinations for every key one has a
single scale, which is a set of basic ratios.  For example, instead of
memorizing major/minor tonic triads for every key, such as "C,E,G",
"C,$^\flat$E,G" for major/minor triads in C-key, etc.  (a total of
2$\cdot$12=24 combinations), one will have to remember only two combinations of
ratios: (5/4, 3/2) for a major triad and (6/5, 3/2) for a minor triad, the
first number in the triad always being 1.  Shifting to different chords in a
progression will amount to multiplication of these numbers by a rational number
selected from the currently used scale.  For example, a sequence (1, 9/8, 5/4,
4/3, 3/2, 5/6, 15/8, 2) will resemble the chromatic major scale.

We will refer to an
octave setup according to the rational scale as {\em instrument scale}.  In this case a harmonic
progression can be replaced by a {\em transposition sequence}, which we will
also refer to as a {\em harmonic sequence} for clarity.  This harmonic
sequence can be further generalized to a hierarchy of harmonic sequences
of different levels.  A harmonic sequence on each level of hierarchy is
derived from a {\em transposition scale}, which is also a rational scale, i.e.
a subset of rational numbers.  Each transposition scale can be equal to the
instrument scale, or can differ from it, thus creating a {\em multi-scale
composition}.  Each harmonic sequence can be related to its own transposition scale, or
a single scale can be used for all levels.  Frequencies of sounds are obtained
by multiplying the base frequency by the factor corresponding the current
instrument key selected from the instrument scale and by a cumulative product
of factors selected from harmonic sequences on all levels and
corresponding to the current time interval in the composition.

One can also extend this method of multi-scale composition to a {\em
multi-scale orchestra} of instruments.  In this case when composing for an
orchestra, harmonic sequences and scales can be assigned independently to
each musical instrument.  This method also opens up opportunities of exploring
different musical scales that can exist within the realm of physical
resonances.

\section{\label{hs}Harmonic Sequences}

In the proposed system of generating musical sounds the fundamental
frequency of each sound is obtained as a product between the base
frequency and a cumulative product of rational multipliers. Looking from a
slightly different perspective, the fundamental frequency is obtained as
a multiple of another frequency, which in turn can be obtained as a
multiple of yet another frequency, and so on. The multipliers can be
selected from a subset of rational numbers, defined by simple ratios of
two integers. 

In the simplest case, which we shall refer to as a {\em level-1} composition, the
procedure of creating a composition starts with a single {\em base frequency}
and a subset of rational numbers, further referred to as the {\em instrument
scale}, and each number in the set will be referred to as the {\em instrument
key}.  Each sound in a composition is characterized by its fundamental
frequency, further referred to as the {\em note frequency}, and the associated
time interval. The note frequency is obtained by multiplying the base frequency
by a factor taken from the instrument scale and corresponding to the current
instrument key. We will refer to this number as the {\em scale key}. 
The pair consisting of the scale key and the corresponding time
interval will be referred to as the {\em note}. 
Thus, the procedure of level-1 composition sets for each time
interval in a composition a corresponding note frequency equal to a product of
the base frequency and one of the keys selected from the instrument scale.
This key selection can be done independently for each instrument.  The time
intervals can be overlapping, thus allowing for playing chords.  This sequence
of notes selected from the instrument scale will be referred to as the {\em
instrument score}. The manner in which the instrument score is compiled is not
important for this discussion and is presumably done by a composer.

\begin{figure}[h]
\centering
\fbox{\includegraphics[width=4.6in]{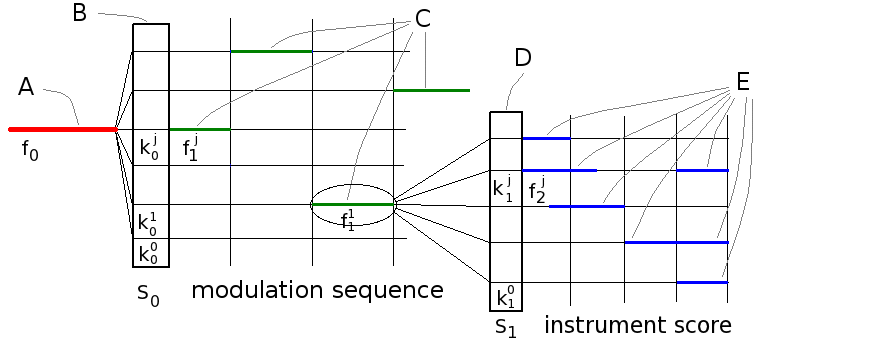}}
\caption{\label{scales}~Assigning frequencies in level-2
composition: 
A - base frequency, 
B - transposition scale, 
C - harmonic sequence, 
D - instrument scale, 
E - instrument score. 
}
\end{figure}

In the next step this procedure is extended to {\em level-2}
composition (Fig.\ref{scales}). In this case in addition to the base
frequency (A), instrument scale (D),
and instrument score (E), there is another layer of
rational numbers further referred to as the {\em transposition scale}
(B), inserted between the base frequency and the
instrument scale.  Similarly to the level-1 composition, a sequence of
numbers is selected from this scale and assigned to the corresponding time
intervals, thus resulting in another sequence of key-time pairs, which
we will refer to as the {\em harmonic sequence} (C), and each pair in the
sequence will be referred to as the {\em transposition tone}.  Unlike the notes of the
instrument score, time intervals of transposition tones should be
non-overlapping and span the entire composition.
Thus, in the notation of Fig.\ref{scales}  
$f_i^j$ represents the frequency 
on $i$-th level, obtained from $j$-th key 
in $i$-th scale, $k_i^j$.

Selection of transposition keys and associated time intervals to make up a
harmonic sequence is done by a composer and it will determine the
harmonic structure of the composition.

As an example consider the level-2 composition. Here the instrument sound frequencies ($f_2^j$) are determined as
triple products of the base frequency, transposition tones, and instrument
keys.  The selection of a transposition tone for each note is always
possible and unique, because the time intervals of transposition tones are
restricted to be non-overlapping and span the entire composition.
In fact, the level-2 composition procedure corresponds to a conventional composition
where the instrument score takes the place of a conventional
score and harmonic sequence takes place of a chord progression
(harmony of the song). 
The difference here is in simplification in the
instrument score due to the effect of frequency shift provided by transposition
sequencing and simpler octave structure as discussed above.

One can further generalize the above procedure to {\em level-n}
composition through an {\em n}-step recursive frequency transformation which is
specified for each time, $t$, in the composition, as: 

\begin{equation}
f_{n+1}(t) = f_n(t)m_n(t) = f_0\prod_{i=0}^nm_i(t)
\label{fn}
\end{equation}

\noindent
where the initial frequency, $f_0$, or the {\em base frequency} will be
a time-constant: $f_0(t)=f_0=const$, and factors $m_n(t)$ for $n>0$,
correspond to transposition tones for time $t$ of a harmonic sequence of level $n$, and
$m_0(t)$ corresponds to the instrument score. Each $m_n(t)$ time sequence
is selected from the corresponding $n$-level scale
$S_n$ as: 

\begin{equation}
m_n(t) = {\hat{\cal R}}_n^{(t)}S_n
\label{kn}
\end{equation}

\noindent
where 
${\hat{\cal R}}_n^{(t)}$ denotes a generally time-dependent selection
operator provided by a composer or an algorithm, and
scale $S_n$ on level $n$ is an ordered
collection of keys, $k_n^i$, represented by a subset of 
rational numbers,
i.e. 

$$S_n = \mathop{\bigcup}_{j}k_n^j$$

The above requirement of non-overlapping time intervals in a harmonic
sequence means that only the $0$-level selection
operator, ${\hat{\cal R}}_{0}^{(t)}$, acting on the instrument
scale in (\ref{kn}), is allowed to generate multiple selections for the same value of $t$,
thereby enabling chords in the instrument score. Chords have no obvious meaning
for a harmonic sequence as defined here.

Let's now consider an example of level-3 composition. Here
we introduce a second
harmonic sequence and optionally the associated transposition
scale. The frequency of sound in this case will be determined as a product of
the base frequency, the key from the currently played note of a particular instrument
score, and the transposition tones from the harmonic sequences of levels 1 and 2, both corresponding to the current time interval.
Thus, according to (\ref{fn}) the key note frequency in an instrument score will be determined as:

\[
f_3 = f_2m_2 = f_1m_1m_2 = f_0m_0m_1m_2
\]


In this case the instrument score is provided by sequence $m_0$, which
corresponds to a conventional score, harmonic sequence $m_1$ will
correspond to a harmonic structure of a song, such as a sequence of keys
assigned to different measures, e.g.  C, Am, F, G, and harmonic sequence
$m_2$ can be related to a conventional transposition, which in this case will
up/down-shift all frequencies in certain parts of the song.  The distinction
from the conventional composition will be in simplification in the
instrument score, because the transition between different keys
(chord progression) is already taken care of by the harmonic sequences. 
Another simplification comes from the lower
cardinality of the octave as well as from the fact that the
problem of selecting appropriate tones corresponding to a current melodic key
no longer exists.

It should be noted that one can also use the reverse definition of sequences
$m_i$, where the last sequence takes place of an instrument score and the rest
perform the transposition, which corresponds to the illustration in
Fig.\ref{scales}.

In the case of multiple instruments (orchestra) it is possible to
assign different instrument scales to different instruments and even
to direct them to follow different harmonic sequences.  In the
simplest case the base-level harmonic sequence may consist of a
single tone, which will correspond to a transposition of that
instrument to a different frequency range for a duration of the
composition.

Even though in the proposed formalism one can use different transposition scales on
different levels in reality it
would seem most practical to use just one transposition scale on all levels.
Also, for human composers and performers it will probably be not very
practical to go beyond level-3 composition.  Nevertheless, higher level
multi-scale compositions can be explored in algorithmically generated
compositions.

In summary, the method of dynamic transposition described here augments the
conventional notions of transposition with the concepts of harmonic sequences
and harmonic scales and the usage of rational numbers to relate note
frequencies to the base frequency through harmonic tones.

\section{Implementation}

\subsection{Music Sequencing}

The principles outlined above can be easily implemented in a
music software by adapting standard
MIDI-sequencer techniques. 
For the purpose of this work an open source LMMS software  
\cite{David:2012:LCG:2432363} was installed on a Linux PC and modified to enable a level-3 composition as described in Sec.\ref{hs}.
The new customized sequencer was used for manual composition in just intonation scale \cite{Tectral15}.

\begin{figure}[h]
\centering
\fbox{\includegraphics[width=4.0in]{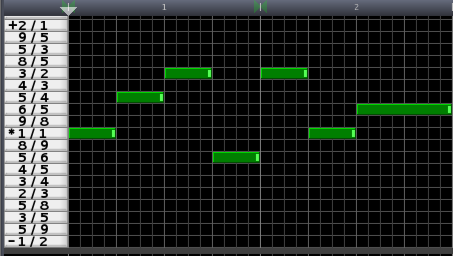}}
\caption{\label{keyboard}~Instrument score or harmonic sequence in a "piano"-roll.
}
\end{figure}

\begin{figure}[h]
\centering
\fbox{\includegraphics[width=4.0in]{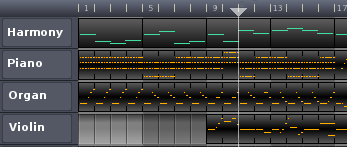}}
\caption{\label{track}~Transposition sequence displayed as a track in a song editor.
}
\end{figure}

Fig.\ref{keyboard} shows a generic
piano-roll type panel where piano keys are replaced with an ordered set
of ratios, which can represent either instrument or transposition scales.
This setup can be used for entering instrument scores as well as
harmonic sequences - all in a similar manner. It should be noted, that
the two octaves of numerical keys shown are all the keys needed
to play an instrument. Shifting to higher or lower octaves can be done by
harmonic sequencing.

Fig.\ref{track} shows a generic view of a song editor, where along with
traditional instrument tracks a harmony track is added. In this case it
provides a global harmonic sequence for all instruments.

\subsection{Instruments}

It is also feasible to design electronic musical instruments based on
these principles.  If implemented in a keyboard fashion, such
instrument will require a small set of keys for a single octave or a
couple of octaves only.  The keys will correspond to a pre-defined
instrument scale. A common 7-tone scale will suffice for most purposes,
resulting in a 7-key keyboard. But even simpler 4-6 key instruments could
be useful for playing basses or for less sophisticated game oriented
devices.  A transposition control for such instrument can be implemented
in a separate key, knob, touch-pad, or similar. 

\begin{figure}[h]
\centering
\subfloat["Soundgun" with a transposition slider and dynamic transposition buttons.]{\fbox{\includegraphics[width=4.8in]{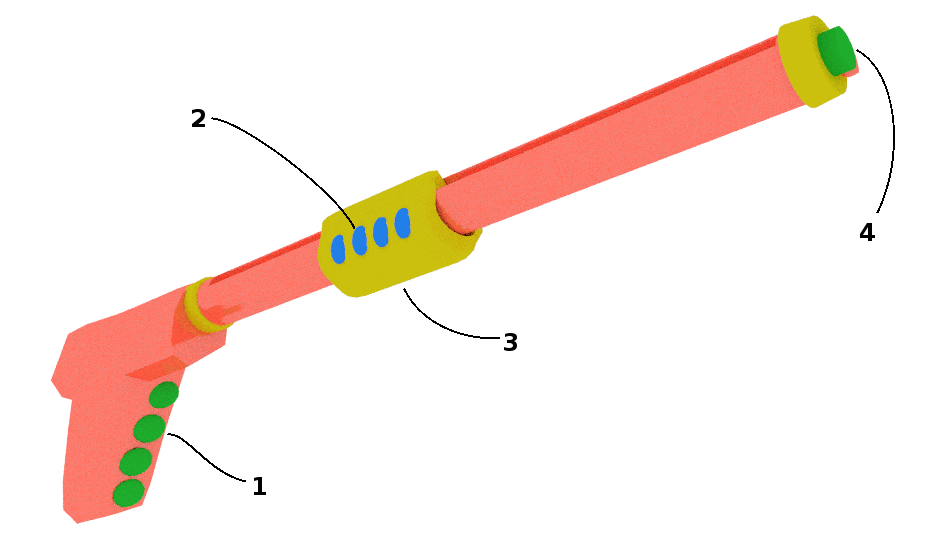}}}\\
\subfloat["Keytar" with dynamic transposition buttons.]{\fbox{\includegraphics[width=4.8in]{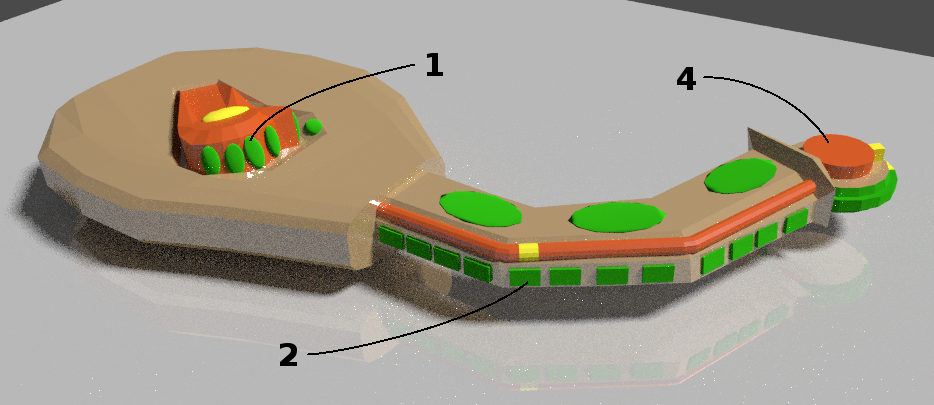}}}
\caption{\label{keytar}~Model instruments based on dynamic transposition. \\
(1) play buttons, (2) transposition buttons, (3) transposition slider, (4) transposition knob.
}
\end{figure}

Figure~\ref{keytar} provides two examples of possible instruments using dynamic transposition.
An even simpler implementation can be based on two joysticks each equipped
with several buttons. In this case one joystick can be used for playing the
musical tones and another for dynamic transposition. 

Whole harmonic
sequences can be programmed-in for the purposes of performance.
In this case switching between transposition tones in a sequence can
be done automatically, or by pushing a separate key.

\subsection{Notation}

No attempt is made in this work to introduce a musical notation to accompany
the proposed method. As long as the technique is not used for life performance
no need for any specific notation arises. Indeed, in the current implementation
all instrument scores and harmonic sequences are recorded and modified by
means of a piano roll editor shown in Fig.\ref{keyboard} and then stored in
respective computer files. The final result is an audio file used for playback,
while the human-readable scores are available either via editor interface of
the respective sequencer application or in XML-encoded text files. The
notation used in those files is similar to the MIDI standard modified for
handling non-chromatic frequencies and harmonic sequences.

Nevertheless, a variety of specific musical notations can be developed for the
purpose of life performance. This can be a simplification of the conventional
notation. For example, the accidentals can be dropped since the cardinality of
the octave is equal to the number of tones in the scale. Thus, all notes can be
labeled by plain letters, including those of harmonic sequences. Numerical
labeling by respective rational multiplies can be used for rigour or
color labeling can be employed to enhance score readability in life performances.
Ultimately, the notation used should depend on the key layout of the
instrument so as to facilitate reading of the score.  Thus, with color labeled
keys the respective color labeling for the score would be appropriate.

Also, to simplify the performance, the harmonic sequence(s) can be stored and
read by the instrument, while the performer will only read the instrument
score. In this case each change of harmonic key in chord progression can be
done either automatically by the instrument or manually by the performer, using
a single key or a foot pedal.

\section{Conclusions}

As mentioned in earlier, the limitation of a just intonation
scale is in its inability to achieve transpositional invariance.  In the
proposed method this limitation is overcome by introducing a generalized
transposition as a system of multiple scales, and hierarchical harmonic
sequences derived from the sets of rational numbers. In this framework it is
now possible to replace harmonic progression with transposition of the base
frequency to any value, and do so independently for different instruments. In
this way one can play the same melodic sequence of tones in different keys and
still realize a simple rational scaling of the base frequency as well as to
retain the uniformity of key patterns within any harmonic key, or indeed within
any frequency range.  Thus, the proposed method enables the usage of musical
scales beyond chromatic, including just intonation and other rational scales.
From the perspective of physical reality of resonances and wave harmonics,
frequencies produced as rational multiples of the fundamental frequency are
more natural, and therefore tend to be more pleasant to human ear, which is
indeed confirmed by old traditions and modern research.

The proposed method can be used for both composing music by means of sequencer
applications and for life performances. It is mostly applicable to electronic
instruments, computers, and other sound-capable digital devices.  In addition
to a more harmonious sound generation the method can simplify writing of
musical scores as well as playing instruments built on these principles. The
method also opens the possibility of exploring multi-level transposition 
technique in algorithmic compositions.


\bibliographystyle{plain}
\bibliography{paper}

\begin{thebibliography}{10}

\bibitem{SpectralMusic01}
Julian Anderson.
\newblock Spectral music.
\newblock In Stanley Sadie and John Tyrrell, editors, {\em The New Grove
  Dictionary of Music and Musicians. Second edition}. Macmillan Publishers,
  2001.

\bibitem{BnMTP2003}
Bruce Benward and Marilyn~Nadine Saker.
\newblock {\em Music: In Theory and Practice}, volume~1.
\newblock McGraw-Hill, 2003.

\bibitem{CarlosTAC87}
W.~Carlos.
\newblock Tuning: At the crossroads.
\newblock {\em Computer Music Journal}, pages 29--43, 1987.

\bibitem{FMsynth73}
J.~Chowning.
\newblock "the synthesis of complex audio spectra by means of frequency
  modulation".
\newblock {\em Journal of the Audio Engineering Society}, 21(7), 1973.

\bibitem{CkMCCS2001}
Perry~R. Cook.
\newblock {\em Music, Cognition, and Computerized Sound: An Introduction to
  Psychoacoustics}.
\newblock MIT Press, 2001.

\bibitem{David:2012:LCG:2432363}
Earl David.
\newblock {\em LMMS: A Complete Guide to Dance Music Production}.
\newblock Packt Publishing, 2012.
\newblock \href{https://lmms.io}{lmms.io}.

\bibitem{Fokker67}
A.D. Fokker.
\newblock On the expansion of the musician's realm of harmony.
\newblock {\em Acta musicologica}, 38:197--202, 1967.

\bibitem{Fokker69}
A.D. Fokker.
\newblock Unison vectors and periodicity blocks in the three-dimensional
  (3-5-7-) harmonic lattice of notes.
\newblock In {\em Proceedings of the KNAW}, volume~72 of {\em B}, pages
  153--168, 1969.

\bibitem{Fontville91}
John Fonville.
\newblock Ben johnston's extended just intonation: A guide for interpreters.
\newblock {\em Perspectives of New Music}, 29(2):121, 1991.

\bibitem{HarryParth}
Bob Gilmore.
\newblock {\em "Harry Partch: A Biography"}.
\newblock Yale University Press, 1998.

\bibitem{Hermode11}
\href{http://www.hermode.com}{Hermode Tuning}.
\newblock {Program Controlled Tuning}.
\newblock \href{http://www.hermode.com}{www.hermode.com}, 2011.

\bibitem{BenJonson06}
B.~Johnston and B.~Gilmore.
\newblock {\em "Maximum Clarity" and Other Writings on Music}.
\newblock Urbana: University of Illinois Press, 2006.

\bibitem{LevYBM2007}
D.~J. Levitin.
\newblock {\em This is your brain on music. The science of human obsession}.
\newblock Plume, 2007.

\bibitem{MlIFATC07}
A.~Milne, W.A. Sethares, and J.~Plamondon.
\newblock Invariant fingerings across a tuning continuum.
\newblock {\em Computer Music Journal}, 31:4:15--32, 2007.

\bibitem{SetharesRTT92}
William~A. Sethares.
\newblock Relating tuning and timbre.
\newblock {\em Experimental Musical Instruments}, IX (2), 1992.

\bibitem{SetharesTTSS2005}
William~A. Sethares.
\newblock {\em Tuning, Timbre, Spectrum, Scale}.
\newblock Springer, 2005.

\bibitem{SmirnovPat2013}
A.V. Smirnov.
\newblock Music machine.
\newblock US Patent 8,541,677, 2013.

\bibitem{Tectral15}
Tectral.
\newblock {Music on a Rational Scale}.
\newblock \href{http://music-machine.appspot.com/music}{tectral.com}, 2015.

\bibitem{WdPM1980}
Alexander Wood and J.~M. Bowsher.
\newblock {\em The physics of music}.
\newblock Greenwood Press, 1980.

\end{thebibliography}

\end{document}